\documentclass[prl,twocolumn,floatfix]{revtex4}

\usepackage{amsmath,amssymb,amsfonts,xspace}
\usepackage{hyperref}

\def\varpi{t}

\def\({\left(}
\def\){\right)}
\def\[{\left[}
\def\]{\right]}
\def\hf{{1\over 2}}

\newcommand{\de}{\mathrm{d}}

\newcommand{\I}{\mathrm{i}}

\newcommand{\cL}{\mathcal{L}}

\newcommand{\ep}{\varepsilon}

\newcommand{\p}{\partial}

\newcommand{\cM}{\mathcal{M}}

\newcommand{\cX}{\mathcal{X}}

\DeclareSymbolFont{AMSa}{U}{msa}{m}{n}
\DeclareSymbolFont{AMSb}{U}{msb}{m}{n}
\DeclareMathSymbol{\fieldR}{\mathalpha}{AMSb}{"52}

\newcommand{\cZ}{\mathcal{Z}}

\newcommand{\cO}{\mathcal{O}}

\newcommand{\cU}{\mathcal{U}}

\newcommand{\nn}{\nonumber}

\newcommand{\IR}{\mathbb{R}}
\newcommand{\IC}{\mathbb{C}}
\newcommand{\IZ}{\mathbb{Z}}

\newcommand{\txi}{\tilde\xi}

\newcommand{\CP}{\IC P^1}

\def\bea{\begin{eqnarray}}
\def\eea{\end{eqnarray}}
\def\be{\begin{equation}}
\def\ee{\end{equation}}
\def\ba{\begin{align}}
\def\ea{\end{align}}
\def\bse{\begin{subequations}}
\def\ese{\end{subequations}}

\def\ba{\bar a}

\def\hgam{\hat \gamma}

\def\hH{h}

\def\cij#1{c}
\def\ci#1{c}

\def\ui#1{^{[#1]}}

\def\txii#1{{\tilde\xi}^{[#1]}}

\def\ai#1{{\alpha}^{[#1]}}

\def\xii#1{\xi_{[#1]}}

\def\Hij#1{H^{[#1]}}

\def\XXint#1#2#3{{\setbox0=\hbox{$#1{#2#3}{\int}$}
\vcenter{\hbox{$#2#3$}}\kern-.5\wd0}}

\def\hHij#1{\hH^{[#1]}}

\def\Fcl{F^{\rm cl}}

\def\gamD#1{\tilde\gamma}

\def\gl#1{{\rm g}}

\def\tt{\rm}

\begin{document}

\title{
{\bf\Large
Fivebrane instantons in Calabi-Yau compactifications
} }

\author{Sergei Alexandrov and Sibasish Banerjee}
\email{salexand@univ-montp2.fr, sibasishbanerjee@live.in}
\affiliation{Laboratoire Charles Coulomb, CNRS UMR 5221,
Universit\'e Montpellier 2, F-34095 Montpellier, France}

\begin{abstract}
We provide the last missing piece of the complete non-perturbative description of the low energy effective action
emerging from Calabi-Yau compactifications of type II string theory --- NS5-brane instanton corrections
to the hypermultiplet moduli space $\cM_H$. We find them using S-duality symmetry of the type IIB formulation. The result is encoded
in a set of holomorphic functions on the twistor space of $\cM_H$ and includes all orders of the instanton expansion.
\end{abstract}

\maketitle

  \section{Introduction}

One of the outstanding problems in string theory is to derive the effective theories
appearing as a low energy limit of string compactifications. 
This is an important step in connecting string theory to the real world,
which can also shed light on its non-perturbative structure.
Eventually, we are interested in compactifications preserving $N=1$
or no supersymmetry, since these cases are relevant from the phenomenological point of view.
However, at present the full low energy description of such compactifications seems to be beyond our reach.

At the same time, compactifications preserving eight supercharges in four dimensions seem to be more tractable,
and still involving very non-trivial and rich physics.
One type of such compactifications is provided by type II string theory on a Calabi-Yau manifold.
During recent years a significant progress has been achieved in getting its non-perturbative
description (see \cite{Alexandrov:2011va,Alexandrov:2013yva} for reviews).
The corresponding effective theory is completely determined by the metric on its moduli space, 
which is a direct product of vector and hypermultiplet components, $\cM_V$ and $\cM_H$.
It is in the latter space where all complications are hidden. Being a quaternion-K\"ahler (QK) manifold \cite{Bagger:1983tt}, 
$\cM_H$ receives instanton corrections coming from branes wrapping non-trivial cycles of the Calabi-Yau \cite{Becker:1995kb}.
There are two types of such branes which contribute to the non-perturbative metric: D-branes and NS5-branes.
Using twistorial techniques \cite{Hitchin:1986ea,MR1327157,Alexandrov:2008nk}, 
which provide a very efficient parametrization of quaternionic manifolds,
the D-brane instantons have been incorporated in a series of works \cite{RoblesLlana:2006is,Alexandrov:2008gh,Alexandrov:2009zh}.
This result left NS5-brane instantons as the only remaining unknown piece of the full non-perturbative picture.

An attempt to include these corrections has been done in \cite{Alexandrov:2010ca}. It was based on the fact 
that in the type IIB formulation S-duality maps D5-branes into NS5-branes, which makes possible to get the latter
once we know the former. However, due to a complicated action of S-duality on the twistor space,
where the D-instantons have the most simple formulation,
this idea had been realized only in the one-instanton approximation.

In this paper we go beyond this restriction and provide a complete description of fivebrane instantons 
which includes all orders of the instanton expansion.
The clue to such a result is an improved parametrization of contact transformations which encode the geometry 
of the twistor space of $\cM_H$. It allows essentially to linearize the action of S-duality and thus to extract
fivebrane instanton corrections to the hypermultiplet metric.
In a companion paper \cite{AB:2014} we will show the consistency of our result with the U-duality symmetry group
of $\cM_H$, which requires an improved realization of these symmetries, and extend it to include contributions from
interactions between fivebrane and D1-D(-1)-instantons.

  \section{Twistor approach and contact geometry}

\subsection{Darboux coordinates and transition functions}

We start with a brief review of twistorial techniques which are necessary tools for describing instanton 
corrections to the metric on $\cM_H$ \cite{Alexandrov:2011va}. The need for these techniques stems from the fact
that a generic QK manifold is not even a complex manifold so that the constraints of the QK geometry 
appear to be highly non-trivial. At the same time, $4n$-dimensional QK spaces $\cM$ are in one-to-one correspondence 
with $(4n+2)$-dimensional K\"ahler spaces $\cZ_\cM$ carrying a {\it holomorphic contact structure}. The latter are known as
{\it twistor spaces} and appear as $\CP$ bundles over the original QK manifolds. 

The presence of the complex and holomorphic contact structures makes the twistor spaces much easier to work with.
In particular, the contact structure can be represented by a set of holomorphic
one-forms $\cX^{[i]}$ on each of the patches of the covering $\cZ_\cM=\cup \cU_i$. They are defined
up to rescalings by nowhere vanishing holomorphic smooth functions, and
such that $\cX^{[i]} \wedge \(\de\cX^{[i]}\)^n \ne 0$ is a holomorphic top form.
Given these one-forms, in each patch one can introduce a system of Darboux coordinates 
$(\xii{i}^\Lambda,\txii{i}_\Lambda,\ai{i})$, $\Lambda=0,\dots,n-1$,
fixed (non-uniquely) by the requirement that
\be
\cX\ui{i} = \de\ai{i} + \xii{i}^\Lambda \de\txii{i}_\Lambda.
\ee
Then all information about the geometry of $\cZ_\cM$, and hence $\cM$, 
is contained in the changes of the Darboux coordinates between different patches. They should preserve 
the contact one-form up to an overall holomorphic factor $\cX\ui{i} =  \hat f_{ij}^{2} \, \cX\ui{j}$.
Such contact transformations can be parametrized by holomorphic functions $\Hij{ij}$ which, by analogy with
canonical transformations, are taken to depend on $\xi^\Lambda$ in patch $\cU_i$ and $\txi_\Lambda, \alpha$ in patch $\cU_j$.
In these terms the gluing conditions read \cite{Alexandrov:2009zh}
\be
\begin{split}
\xii{j}^\Lambda = &\, \xii{i}^\Lambda  -\p_{\txii{j}_\Lambda }\Hij{ij}
+\xii{j}^\Lambda \, \p_{\ai{j} }\Hij{ij},
\\
\txii{j}_\Lambda =&\,  \txii{i}_\Lambda
 + \p_{\xii{i}^\Lambda } \Hij{ij}  ,
\\
\ai{j} =&\,   \ai{i}
 + \Hij{ij}- \xii{i}^\Lambda \p_{\xii{i}^\Lambda}\Hij{ij} ,
\end{split}
\label{glucon}
\ee
which results in $\hat f_{ij}^2=1-\p_{\ai{j} }\Hij{ij}$.
Supplemented by proper regularity conditions, these relations can be rewritten as integral equations and solved
with respect to Darboux coordinates as functions of coordinates on the QK base and the $\CP$ fiber.
Starting from these solutions, there is a straightforward, although a bit non-trivial procedure to derive the metric 
on $\cM$ \cite{Alexandrov:2008nk}.

Thus, the twistorial description encodes the geometry of a QK space into a covering of its twistor space
and the associated set of holomorphic functions $\Hij{ij}$. In fact, for the purpose of constructing the {\it local} metric on $\cM$,
the recipe can even be simplified: it is sufficient to provide a set of contours $\ell_i$ on $\CP$ and a set of transition functions $\Hij{i}$
attached to each contour.
Whereas the closed contours typically correspond to boundaries of open patches $\cU_i$, open contours arise as an effective description 
of transition functions with branch cuts. We refer to \cite{Alexandrov:2011va} for more details.

Applying this approach to the problem of computing the non-perturbative metric on the hypermultiplet moduli space $\cM_H$, 
we see that it reduces to the problem of finding contours and holomorphic functions for each type of quantum corrections
contributing to the metric. The perturbative metric was put into this language in \cite{Alexandrov:2008nk} and the twistor data
for D-instantons have been found in \cite{Alexandrov:2008gh,Alexandrov:2009zh}. 
As a result, a D-instanton of charge $\gamma=(p^\Lambda,q_\Lambda)$
is generated by the data consisting of the contour 
\be
\ell_\gamma=\{\varpi\in \CP:\ Z_\gamma/t\in\I\IR^-\},
\ee
where $ Z_\gamma$ is the central charge of the $N=1$ supersymmetry algebra preserved by the D-brane,
and the transition function
\be
\Hij{\gamma} = H_{\gamma} - \frac12\, q_{\Lambda}p^\Lambda  \(H'_{\gamma}\)^2,
\label{trHij}
\ee
where $H_{\gamma}$ is given by
\be
H_{\gamma}(\Xi_{\gamma})
=  \frac{\bar \Omega(\gamma)}{4 \pi^2}\, \sigma_D (\gamma) \, e^{-2\pi\I \Xi_{\gamma}}.
\label{Hgam}
\ee
Here $\sigma_D (\gamma)$ is the the so-called quadratic refinement, 
$\bar \Omega(\gamma)$ are rational DT invariants \footnote{The original result has been formulated in terms of the di-logarithm and
integer DT invariants, which also makes explicit its relation to Thermodynamic Bethe Ansatz \cite{Gaiotto:2008cd,Alexandrov:2010pp}.
The two formulations are readily equivalent.}, and
we used the notation $\Xi_\gamma=q_\Lambda \xi^\Lambda-p^\Lambda\txi_\Lambda$.

\subsection{Contact bracket and improved transition functions}

Although the parametrization of the contact transformations \eqref{glucon} using transition functions $\Hij{ij}$
is very explicit, it also has some inconveniences. The most important problem comes from that the arguments of $\Hij{ij}$
belong to different patches. As a result, all symmetries of the twistor space are realized on the transition functions 
in a very non-trivial way. An example is the symplectic invariance of the D-instanton corrections: 
whereas the function \eqref{Hgam} is clearly symplectic invariant, 
as $(p^\Lambda, q_\Lambda)$ and $(\xi^\Lambda,\txi_\Lambda)$ transform as vectors under symplectic transformations,
this is not the case for the transition function \eqref{trHij}.
This suggests that there should be another way to parametrize the contact transformations where the fundamental role 
is shifted to the function \eqref{Hgam} \footnote{In fact, such a symplectic invariant description has been already proposed in
\cite{Alexandrov:2009zh}. But it works only if transition functions satisfy a certain integrability condition, 
which is indeed the case for $H_\gamma$, but turns out not to be the case for the NS5-transition functions found below.}.

To introduce the new parametrization, we need first to define the so-called ``contact bracket", 
which can be viewed as a lift of the Poisson bracket to the realm of contact geometry and was defined previously in 
\cite[Eq.(2.44)]{Alexandrov:2008gh}. 
This can be done in a coordinate independent way as follows.
Let us associate with a function $h$   
the ``contact vector field" $X_h$ determined by the following relations
\be
\iota_{X_h}\de \cX=-\de h+ R(h) \cX,
\qquad
\iota_{X_h}\cX=h,
\label{defXh}
\ee
where $\iota_X$ is contraction of vector $X$ with a differential form, 
$\cX$ is the contact one-form, and 
$R$ is the Reeb vector field which is the unique element of the kernel of $\de\cX$ such that $\cX(R) =1$.
Then the contact bracket between two functions $h$ and $f$ is defined as
\be
\{h,f\}=X_h(f).
\ee
In terms of Darboux coordinates, it reads explicitly as
\be
\begin{split}
\{h,\xi^\Lambda\}=&\, -\p_{\txi_\Lambda} h+\xi^\Lambda\p_\alpha h,
\qquad
\{h,\txi_\Lambda\}=\p_{\xi^\Lambda} h,
\\
&\qquad
\{h,\alpha\}=h-\xi^\Lambda\p_{\xi^\Lambda} h.
\end{split}
\label{contbr}
\ee
Note that the bracket is not antisymmetric, but satisfies
\be
\{h,h\} =hR(h)=h\p_\alpha h,
\label{commHH}
\ee
which can be obtained by applying $\iota_{X_h}$ to the first equation in \eqref{defXh}.
Besides, the bracket does not follow the Leibnitz rule in the first argument giving instead
\be
\begin{split}
\{ h_1 h_2,f\}=&\, 
h_1\{h_2,f\}+h_2\{h_1,f\}-h_1 h_2\p_\alpha f.
\end{split}
\label{notLeibn}
\ee
The reason for the different behaviour with respect to the two arguments is that actually 
they are supposed to be local sections of different bundles, $\cO(2)$ and $\cO(0)$, 
respectively, and the contact bracket maps them into an $\cO(0)$ section \cite{Alexandrov:2008gh}.

A crucial property of the contact bracket, which immediately follows from 
its definition, is that it generates an infinitesimal transformation scaling the contact one-form
\be
\cL_{X_h}\cX=\de\(\iota_{X_h}\cX\)+\iota_{X_h}\de\cX=R(h)\cX,
\ee
i.e. a contact transformation. Conversely, any {\it finite} contact transformation, for instance, the one which 
relates Darboux coordinates in different patches, 
can be parametrized by a holomorphic function $\hHij{ij}$
and generated by the contact bracket via the following action
\be
\Xi^{[j]}=e^{\{\hHij{ij},\, \cdot\,\}}\, \Xi^{[i]},
\label{glcond-KS}
\ee
where $\Xi^{[i]}$ denotes the set of Darboux coordinates in the patch $\cU_i$.
What is important here is that $\hHij{ij}$ is considered as a function of Darboux coordinates in {\it one} patch.
We call it {\it improved transition function}.

Comparing the gluing conditions \eqref{glcond-KS} and \eqref{glucon}, one obtains relations between
the two types of transition functions.
In particular, this gives the following explicit expression (we dropped the patch indices)
\be
H=\(e^{\{\hH,\, \cdot\,\}}-1\)\alpha+\xi^\Lambda\( e^{\{\hH,\, \cdot\,\}}-1\)\txi_\Lambda.
\label{relHH}
\ee

To illustrate the new parametrization, let us apply the relation \eqref{relHH} to the D-instanton case taking 
the improved transition function to be $\hH=H_\gamma(\Xi_\gamma)$. Using the fact that
$\alpha$-independent functions commute with themselves (see \eqref{commHH}), one obtains
\bea
\( e^{\{H_\gamma,\, \cdot\,\}}-1\)\txi_\Lambda&=& q_\Lambda H'_\gamma,
\\
\(e^{\{H_\gamma,\, \cdot\,\}}-1\)\alpha&=& H_\gamma-q_\Lambda\xi^\Lambda H'_\gamma-\hf\, p^\Lambda q_\Lambda(H'_\gamma)^2.
\nn
\eea
Substituting this into \eqref{relHH}, one reproduces the previous result \eqref{trHij}.

  \section{S-duality in twistor space}

As we explained in the introduction, we are going to find fivebrane instanton contributions to the metric on $\cM_H$
by applying S-duality to the D5-brane corrections. Therefore, it is important to understand how S-duality acts 
at the level of the twistor space, in particular, on Darboux coordinates and on transition functions. 

This question has been already addressed in the previous works \cite{Alexandrov:2008gh,Alexandrov:2012bu,Alexandrov:2013mha}.   
It was found that for an $SL(2,\IZ)$ transformation $\gl{c,d}=\(\begin{array}{cc}
a & b
\\
c & d
\end{array}\)$ to be an isometry of $\cM_H$ lifted to the twistor space the Darboux coordinates should transform as
\bea
&
\displaystyle{\xi^0 \mapsto \frac{a \xi^0 +b}{c \xi^0 + d} \, , \qquad
\xi^a \mapsto \frac{\xi^a}{c\xi^0+d} \, ,} 
&
\nn\\
&
\displaystyle{\txi_a \mapsto \txi_a +  \frac{c\kappa_{abc} \xi^b \xi^c}{2(c \xi^0+d)}- c_{2,a}\ep(\gl{c,d})\, ,}
&
\label{SL2Zxi}\\
&
\begin{pmatrix} \txi_0 \\ \alpha \end{pmatrix} \!\mapsto\!
\begin{pmatrix} d & -c \\ -b & a  \end{pmatrix}
\begin{pmatrix} \txi_0 \\  \alpha \end{pmatrix}
- \displaystyle{\frac{c}{6}\, \kappa_{abc} \xi^a\xi^b\xi^c}
\!\begin{pmatrix}
\frac{-c}{c \xi^0+d}\\
\frac{c (a\xi^0 + b)+2}{(c \xi^0+d)^2}
\end{pmatrix}\!,
\nn
\eea
where $a=1,\dots,n-1$, 
$\kappa_{abc}$ are triple intersection numbers of the Calabi-Yau,
$c_{2,a}$ are components of the second Chern class along a basis of 2-forms, 
and $\ep(\gl{c,d})$ is the multiplier system of the Dedekind eta function.

Combined with the gluing conditions \eqref{glucon}, the transformation \eqref{SL2Zxi} can be used 
to get the transformation property of the transition functions $\Hij{ij}$.
Although its explicit form was found \cite{Alexandrov:2013mha}, it is highly non-linear and its direct application
is rather involved. We do not give it here since we found a way to proceed which is much more instructive and elegant.
The idea is that the passage to the improved transition functions $\hHij{ij}$ will also improve
transformation properties under S-duality: one can hope that the non-linearities appearing 
in the transformation law for $\Hij{ij}$ are of the same origin as those in \eqref{trHij} and will disappear
when one works in terms of $\hHij{ij}$.

To show that this is indeed the case, let us note the following property of the contact bracket
\be
\{ (c\xi^0+d)\gl{c,d}\cdot h, \gl{c,d}\cdot f\}=\gl{c,d}\cdot \{h,f\},
\ee
which can be verified by direct calculation using \eqref{contbr} and \eqref{SL2Zxi}.
With its help it becomes easy to evaluate the operator relating Darboux coordinates in two patches after the $SL(2,\IZ)$ transformation.
One finds
\be
\begin{split}
\gl{c,d}\cdot e^{\{\hH,\, \cdot\,\}} \cdot\gl{c,d}^{-1}
=&\, 
e^{\gl{c,d}\cdot\{\hH,\, \cdot\,\}\cdot\gl{c,d}^{-1}}
\\
=&\, 
e^{\{(c\xi^0+d)\gl{c,d}\cdot \hH,\, \cdot\,\}}.
\end{split}
\ee
This implies that a QK manifold carries the isometric action of the S-duality group $SL(2,\IZ)$ only 
if the improved transition functions on its twistor space are split into $SL(2,\IZ)$ multiplets and transform {\it linearly }
inside each multiplet with weight $-1$, e.g. \footnote{It is possible also to have on the r.h.s. some regular contributions, which can always be absorbed 
into a redefinition of Darboux coordinates not affecting the contact structure.}
\be
\hH_{m,n}\mapsto \frac{\hH_{m',n'}}{c\xi^0+d},
\quad
\begin{pmatrix} m'\\ n' \end{pmatrix} =
\begin{pmatrix}
a & c
\\
b & d
\end{pmatrix}
\begin{pmatrix} m \\ n \end{pmatrix},
\label{transhH}
\ee
where the pair $(m,n)$ labels the elements of the multiplet.

  \section{Fivebrane instantons}

An important property of type IIB string theory compactified on a Calabi-Yau is 
that quantum corrections to the hypermultiplet moduli space $\cM_H$ are arranged into different sectors
invariant under S-duality. This happens according to the following pattern:
\be
\mbox{
\begin{tabular}{l|c|c|c|c|c|c|cc|}
\cline{2-2} \cline{4-4}
$\alpha'$: \hspace{0.1cm} & perturb. & \hspace{0.1cm} & w.s. inst  & \multicolumn{4}{c}{} \rule{0pt}{10pt}
\\
\cline{6-6} \cline{8-9}
$g_s$: & 1-loop \ D(-1) & & D1 & \hspace{0.1cm} & \,D3\, & \hspace{0.1cm} & \,D5 &  NS5 \rule{0pt}{11pt}
\\
\cline{2-2} \cline{4-4} \cline{6-6} \cline{8-9}
\end{tabular}}
\label{quantcor}
\ee
Thus, D(-1)-instantons are combined with perturbative $\alpha'$ and $g_s$-corrections, 
D1-instantons mix with worldsheet instantons, D3-instantons are S-duality invariant, whereas D5 and NS5-instantons
transform as a doublet under $SL(2,\IZ)$. This splitting allows to study each sector independently of the others.
In particular, a twistorial description of the first two sectors, which is manifestly S-duality invariant, 
has been given in \cite{Alexandrov:2009qq,Alexandrov:2012bu}. 
The sector of D3-branes has been studied in \cite{Alexandrov:2012au},
where it was shown that the transition functions \eqref{trHij} restricted to this sector are consistent with S-duality.
Here we concentrate on the sector of fivebranes and derive all corresponding instanton corrections
from the knowledge of D5-instantons. 

In type IIB theory D5-branes, or more precisely D5-D3-D1-D(-1)-bound states, 
are characterized by the rational valued generalized Mukai vector 
$\gamma=(p^0,p^a,q_a,q_0)$ with $p^0\ne0$ \cite{Douglas:2006jp}.
Below we will also need the so-called invariant charges \cite{Alexandrov:2010ca}
\be
\begin{split}
\hat q_a = &\, \textstyle{q_a  + \frac12 \,\kappa_{abc} \frac{p^b p^c}{p^0},}
\\
\hat q_0 =&\, \textstyle{ q_0  + \frac{p^a q_a}{p^0} + \frac13\, \kappa_{abc}\frac{p^a p^b p^c}{(p^0)^2}}
\end{split}
\ee
and the reduced charge vector $\hgam=(p^a,\hat q_a,\hat q_0)$.

Since fivebrane instantons form a doublet of $SL(2,\IZ)$, their 
improved transition functions must follow the transformation law \eqref{transhH}.
Identifying D5-branes with the component $(m,n)=(0,p^0)$, all of them can be obtained by applying $\gl{c,d}$ 
to the function $\hHij{\hgam}_{0,p^0}=H_\gamma$. Taking into account the physical interpretation of the charges,
it is convenient to take  
\be
c= - k/p^0,
\qquad  
d=p/p^0,
\qquad
p^0=\gcd(p,k).
\ee
Then $k$ appears to be precisely the NS5-brane charge.
Using \eqref{SL2Zxi}, it is straightforward to obtain
\be
\hHij{\hgam}_{k,p}=-\frac{\bar\Omega_{k,p}(\hgam)}{4\pi^2} \frac{k}{p^0} (\xi^0-n^0)\sigma_D(\gamma)\, e^{2\pi\I S_{k,p;\hgam}},
\label{fivebraneh}
\ee
where
\bea
S_{k,p;\hgam}&=& k\(\alpha + n^\Lambda \txi_\Lambda  + \Fcl (\xi - n)\)
\\
&+& \frac{p^0(k \hat q_a (\xi^a - n^a)+ p^0 \hat q_0)}{k^2(\xi^0 - n^0)} + \frac{a}{k}\,p^0 q_0- c_{2,a} p^a \varepsilon(\gl{c,d}),
\nn
\eea
$\Fcl(X)=-\kappa_{abc} \,\frac{X^a X^b X^c}{6 X^0}$ is the classical prepotential, $n^a=p^a/k$, $n^0=p/k$,
and $\bar\Omega_{k,p}(\hgam)\equiv\bar\Omega(\gamma;\gl{c,d}\cdot z)$ takes into account the fact that DT invariants
are only piecewise constant and generically jump along lines of marginal stability in the moduli space of K\"ahler structure deformations $z^a$.
It should not be surprising that the resulting function \eqref{fivebraneh}, 
up to the factor ensuring the correct modular weight, coincides with the result found in \cite[Eq.(5.30)]{Alexandrov:2010ca}
in the one-instanton approximation: in this approximation the two types of transition functions are identical and 
the remarkable feature of \eqref{fivebraneh} is that it is {\it exact} at linear order.

Furthermore, it is possible to evaluate explicitly the contact transformation generated by the function $\hHij{\hgam}_{k,p}$
by applying the operator \eqref{glcond-KS}. Referring to \cite{AB:2014} for details,
here we give the result just for the transition function
defined by the relation \eqref{relHH} \footnote{The function (\ref{tranNS5all}) is written in terms of Darboux coordinates in one patch what is not
sufficient to calculate its derivatives entering the gluing conditions (\ref{glucon}). Their expressions can be found in \cite{AB:2014}.}
\be
\Hij{\hgam}_{k,p}
= \hHij{\hgam}_{k,p}
-2\pi^2(\hHij{\hgam}_{k,p})^2\[\frac{\hat q_0 (p^0)^2}{k(\xi^0-n^0)}-\frac{2k^2\Fcl(\xi-n)}{\(1-2\pi\I k\hHij{\hgam}_{k,p}\)^2} \].
\label{tranNS5all}
\ee
One observes that this function generates an infinite expansion in instantons equivalent to the expansion in DT invariants
or in $\hHij{\hgam}_{k,p}$. In \cite{AB:2014} we also show that it solves the non-linear S-duality constraint
found in \cite{Alexandrov:2013mha} and is consistent with all discrete symmetries generating the U-duality group of the compactified theory.

  \section{Conclusions}

In this paper we found a twistorial description of fivebrane instanton corrections to the hypermultiplet moduli space $\cM_H$ of 
type II string theory on a Calabi-Yau. It is provided by the holomorphic functions \eqref{fivebraneh} and \eqref{tranNS5all}.
With these results at hand, one can now write integral equations for Darboux coordinates on the twistor space of $\cM_H$,
whose solution uniquely defines the metric on the moduli space and thereby the non-perturbative low energy effective action
of the compactified theory.

This progress has become possible due to a new parametrization \eqref{glcond-KS} 
of contact transformations with the help of the so-called contact bracket.
The improved transition functions $\hHij{ij}$ entering this parametrization appear to be more fundamental
than their ordinary cousins $\Hij{ij}$. As a result, all transformation laws and results for instantons 
take a much simpler linear form being reformulated in their terms.

Although our results provide a complete description of the fivebrane sector of quantum corrections to $\cM_H$,
it still remains to put all quantum effects, shown in \eqref{quantcor}, into one unifying picture. 
This problem is addressed in \cite{AB:2014},
where we provide the twistor space construction including all sectors except the one of D3-branes. 
The latter represents a challenge since, despite it is captured by the transition functions \eqref{trHij}, 
even in the one-instanton approximation it has not been reformulated yet in
an explicitly S-duality invariant way. It is expected that a crucial role in such a reformulation
will be played by mock modular forms \cite{Alexandrov:2012au}.

Another interesting problem is to put our type IIB construction into the mirror type IIA formulation.
In particular, it is interesting to see how the NS5-brane instantons deform the integrable structure 
of the D-instantons encoded in the Thermodynamic Bethe Ansatz equations \cite{Gaiotto:2008cd,Alexandrov:2010pp}.

Besides, there is a number of important issues which can be approached once the fivebrane instantons have been incorporated.
These include: a resolution of the singularity generated by the one-loop correction on $\cM_H$;
divergence of the sum over charges due to the exponential growth of DT invariants in supergravity, 
which was argued to be related to the NS5-brane effects \cite{Pioline:2009ia}; 
and consistency of the NS5-brane instantons with wall-crossing.

Finally, one can hope that the results presented here can be useful as well for phenomenological studies.
For instance, in \cite{Ketov:2014roa} it was argued that the fivebrane instantons may be crucial for the derivation 
of the Starobinsky model \cite{Starobinsky:1980te} of the inflationary cosmology from compactifications of string theory.

\section*{Acknowledgments}

We are grateful to Sylvain Ribault for careful reading of the manuscript.


\providecommand{\href}[2]{#2}\begingroup\raggedright\endgroup

\end{document}